\documentclass{article}
\usepackage{graphicx} 
\usepackage{color}
\usepackage{geometry}
\usepackage[size=small]{todonotes}
\usepackage{multirow}
\usepackage{booktabs}
\usepackage{float}
\usepackage{tabularx}
\usepackage{adjustbox}
\usepackage{xcolor}
\usepackage{hyperref}
\usepackage{array}
\usepackage{amsmath,amssymb}
\usepackage{csvsimple} 
\usepackage{hyperref}
\usepackage{xcolor}
\usepackage{longtable} 

\newcommand{\arxivlarge}[0]{\texttt{arxiv-0623}}
\newcommand{\arxivsmall}[0]{\texttt{arxiv-0623-top40}}

\title{
NLLG Quarterly arXiv Report 06/23: \\
What are the most influential current 
AI 
Papers?
}
\author{Steffen Eger, Christoph Leiter, Jonas Belouadi \\Ran Zhang, Aida Kostikova, Daniil Larionov, Yanran Chen, Vivian Fresen\\ \texttt{Natural Language Learning Group (NLLG)}, \url{https://nl2g.github.io/}}
\date{}

\begin{document}

\maketitle

\begin{abstract}
The rapid growth of information in the field of Generative Artificial Intelligence (AI), particularly in the subfields of Natural Language Processing (NLP) and Machine Learning (ML), presents a significant challenge for researchers and practitioners to keep pace with the latest developments. To address the problem of information overload, this report by the Natural Language Learning Group at Bielefeld University focuses on identifying the most popular papers on arXiv, with a specific emphasis on NLP and ML. The objective is to offer a quick guide to the most relevant and widely discussed research, aiding both newcomers and established researchers in staying abreast of current trends. In particular, we compile a list of the 40 most popular papers based on normalized citation counts from the first half of 2023. We observe the dominance of papers related to Large Language Models (LLMs) and specifically ChatGPT during the first half of 2023, with the latter showing signs of declining popularity more recently, however. Further, NLP related papers are the most influential (around 60\% of top papers) even though there are twice as many ML related papers in our data. Core issues investigated in the most heavily cited papers are:  LLM efficiency, evaluation techniques, ethical considerations, embodied agents, and problem-solving with LLMs. 
Additionally, we examine the characteristics of top papers in comparison to others outside the top-40 list (noticing the top paper's focus on LLM related issues and higher number of co-authors) and analyze the citation distributions in our dataset, among others. 

\end{abstract}

\section{Introduction}


In an era of ever-accelerating information flow, staying abreast of the overwhelming flood of data and research output 
is an intimidating task. 
This holds true especially in the context of the current 
large public interest (and even hype) surrounding Generative AI, with papers disseminated in ever shorter time intervals. 
This report, published by the Natural Language Learning Group (\url{https://nl2g.github.io/}) at Bielefeld University, aims to 
alleviate the information overload problem, even if only by a small extent, by identifying  
the currently most popular papers on the arXiv (\url{https://arxiv.org/}), especially focusing on the AI subfields natural language processing (NLP) and machine learning (ML) as some of the most vividly discussed research areas, including in mainstream media. 
Our intention is to give practitioners, incomers, and users of AI, 
from related and non-related fields (e.g., the social sciences or digital humanities) a quick guide on the most popular and presumably most relevant papers 
in order to better and (more) quickly grasp current developments. 

We place particular emphasis on exploring arXiv,\footnote{Our report is similar to a `conference report' as a popular form of science communication, e.g., \url{https://www.romanklinger.de/blog-assets/2023-05-12/eacl2023-conf-report.pdf}. But instead of focusing on conferences, we focus on arXiv for multiple reasons: among others, (i) in an age of rapid developments, conferences and journals are too slow and often lagging behind recent developments; (ii) as everyone who regularly submits to NLP/ML conferences knows, conferences also suffer from low reviewing quality, with junior and non-expert reviewers abounding. Instead, we focus on citations (even though these are not unproblematic themselves) as a form of 
large-scale crowd voting. 
}
given its status as a comprehensive and extremely popular pre-print repository. 
 Notably, arXiv's expedited publication process provides a distinct advantage over traditional conferences and journals, ensuring that the latest 
 research becomes readily available to the scientific community at a much faster pace. 
 
  This report is structured as follows. In Section \ref{sec:methodology}, we outline our methodology, which is entirely straightforward: we select papers from arXiv from the first half of the year 2023 and sort them by normalized citation counts. In Section \ref{sec:toppapers}, we show and discuss the list of the 40 most popular papers --- in terms of normalized citation counts --- from our arXiv dataset. In Section \ref{sec:analysis}, we provide an analysis of our arXiv dataset relating to citation distributions, arXiv categories involved, characteristics of top papers, and popularity of `hype' concepts such as ChatGPT and large language models (LLMs) over time. In Section \ref{sec:conclusion}, we conclude. 

  Among our key findings are that: (i) NLP, once a niche area of research, is now considerably more influential than ML in terms of the citations it attracts: even though there are twice as many ML papers in our datasets, $\sim$60\% of the most highly cited papers are from NLP; (ii) LLM and ChatGPT related papers have clearly dominated the first half of 2023, but especially ChatGPT is now on the decline; (iii) the efficient open-source model LLaMA from Meta AI is the relatively and absolutely most cited paper in our dataset, leaving behind the larger and properietary ChatGPT and GPT-4.  

Our code and data is available from \url{https://github.com/NL2G/Quaterly-Arxiv}.

\section{Methodology}\label{sec:methodology}
 To identify the most influential papers from the AI subfields 
 NLP and ML, 
 we used the following methodology.

\begin{table}[!htbp]
    \centering
    \begin{tabular}{c|ccc}
         \toprule
         \textbf{Dataset name} & \textbf{Size} & \textbf{Time period} & \textbf{\# Primary Categories}\\ \midrule
         \arxivlarge &  20,843 & 01/01/2023-06/31/2023 & 123\\
         \arxivsmall &  40 & 01/01/2023-06/31/2023 & 5\\
         \bottomrule
    \end{tabular}
    \caption{Elementary statistics on our two released datasets. Size is the number of papers in each dataset; the last column gives the number of distinct primary arXiv categories our papers are assigned to.}
    \label{table:stats}
\end{table}

\begin{enumerate}
\item \textbf{Data Retrieval from arXiv}: 
    We collect all papers from 01/01/2023 to 06/31/2023 belonging to the arXiv categories cs.CL (computation and language) 
    and cs.LG (machine learning) using a Python arXiv API.\footnote{ArXiv papers may belong to several categories. We only require that one of the involved categories be one of the two indicated.} 
    \footnote{We did not include cs.AI (without cs.LG or cs.CL) but we note that our top-40 list would have looked very similar with our without the cs.AI requirement. In particular, all top-40 papers would have remained the same --- also note that many cs.AI papers are still included in our dataset, see below.}
    Our retrieval time is \textbf{July, 29, 2023} (which is important, because citation counts are constantly in flux). ArXiv papers can be updated anytime; we take the date of the first submission of a paper to arXiv as its publication date.

\item \textbf{z-score calculation:}
 For each paper, we extract its citation count, as a measure of popularity and arguably importance \cite{doi:10.1177/2158244019829575}, from Semantic Scholar \url{https://www.semanticscholar.org/}. Since papers published at different time points may naturally have different citation counts (e.g., older papers have higher chance of being cited than very novel papers), we calculate a \emph{normalized citation count} by determining \emph{how many standard deviations a paper is above the mean of citations of all papers published in the same week (Sunday-Saturday)}. This is the so-called z-score of Newman \cite{newman2014prediction}:
 \begin{align*}
    z_t = \frac{c_t-\textit{mean}(\mathbf{c}(t))}{\textit{std}(\mathbf{c}(t))}
 \end{align*}
 for a paper published in week $t$ with citation count $c_t$; $\mathbf{c}(t)$ is the list of citation counts of all papers published in week $t$. 
 If a paper lies several standard deviations above the mean (for all papers published in the same week), it can be considered excellent for its class. For example, in a normal distribution, only about 16\% of data points lie one standard deviation above the mean value. As will be seen below, our top papers lie at least 9--12 standard deviations above the mean.\footnote{Our approach of identifying top papers in arXiv via the zscore is similar to \cite{eger2019predicting}.} 

   
\item \textbf{Manual Evaluation}
 The published date on arXiv might differ from the actual first publication/release/submission date of a paper, e.g., when the authors upload the paper much later to arXiv. Thus, we conduct a manual evaluation to verify if a paper 
 genuinely appeared the first time as indicated by its arXiv release time stamp. If the paper was available earlier, we remove it from consideration. 

 


\end{enumerate}

Steps 1 and 2+3 above result in two distinct datasets that we release with this report. We refer to them as \arxivlarge{} and \arxivsmall{}, respectively. Table \ref{table:stats} gives elementary statistics on each of them. 


 \begin{table}[!htbp]
 \centering
 \small 
 \begin{tabular}{|p{0.3cm}|p{5.5cm}|p{1cm}|p{4.5cm}|p{0.7cm}|p{0.5cm}|p{1cm}|}
\toprule
 No. & Title & Cat. & Link & Week & Cit&  z-score \\ \midrule
1 & LLaMA: Open and Efficient Foundation Language Models & cs.CL & \url{http://arxiv.org/abs/2302.13971v1} & 9 & 874 & 28.051\\
2 & GPT-4 Technical Report & cs.CL & \url{http://arxiv.org/abs/2303.08774v3} & 11 & 509 & 25.382\\
3 & PaLM 2 Technical Report & cs.CL & \url{http://arxiv.org/abs/2305.10403v1} & 20 & 82 & 25.182\\
4 & Sparks of Artificial General Intelligence: Early experiments with GPT-4 & cs.CL & \url{http://arxiv.org/abs/2303.12712v5} & 12 & 354 & 24.302\\
5 & PaLM-E: An Embodied Multimodal Language Model & cs.LG & \url{http://arxiv.org/abs/2303.03378v1} & 10 & 164 & 21.225\\
6 & QLoRA: Efficient Finetuning of Quantized LLMs & cs.LG & \url{http://arxiv.org/abs/2305.14314v1} & 21 & 30 & 19.944\\
7 & Segment Anything & cs.CV & \url{http://arxiv.org/abs/2304.02643v1} & 14 & 165 & 18.548\\
8 & Judging LLM-as-a-judge with MT-Bench and Chatbot Arena & cs.CL & \url{http://arxiv.org/abs/2306.05685v2} & 23 & 21 & 17.916\\
9 & A Multitask, Multilingual, Multimodal Evaluation of ChatGPT on Reasoning, Hallucination, and Interactivity & cs.CL & \url{http://arxiv.org/abs/2302.04023v2} & 6 & 214 & 16.819\\
10 & A Survey of Large Language Models & cs.CL & \url{http://arxiv.org/abs/2303.18223v11} & 13 & 169 & 16.594\\
11 & Visual Instruction Tuning & cs.CV & \url{http://arxiv.org/abs/2304.08485v1} & 16 & 89 & 15.277\\
12 & Tree of Thoughts: Deliberate Problem Solving with Large Language Models & cs.CL & \url{http://arxiv.org/abs/2305.10601v1} & 20 & 49 & 14.968\\
13 & Voyager: An Open-Ended Embodied Agent with Large Language Models & cs.AI & \url{http://arxiv.org/abs/2305.16291v1} & 21 & 21 & 13.860\\
14 & Toolformer: Language Models Can Teach Themselves to Use Tools & cs.CL & \url{http://arxiv.org/abs/2302.04761v1} & 6 & 175 & 13.716\\
15 & How Close is ChatGPT to Human Experts? Comparison Corpus, Evaluation, and Detection & cs.CL & \url{http://arxiv.org/abs/2301.07597v1} & 3 & 94 & 13.712\\
16 & Extracting Training Data from Diffusion Models & cs.CR & \url{http://arxiv.org/abs/2301.13188v1} & 5 & 97 & 13.596\\
17 & Large Language Models are not Fair Evaluators & cs.CL & \url{http://arxiv.org/abs/2305.17926v1} & 22 & 14 & 13.352\\
18 & HuggingGPT: Solving AI Tasks with ChatGPT and its Friends in Hugging Face & cs.CL & \url{http://arxiv.org/abs/2303.17580v3} & 13 & 129 & 12.614\\
19 & A Watermark for Large Language Models & cs.LG & \url{http://arxiv.org/abs/2301.10226v3} & 4 & 76
 & 12.481\\
20 & DetectGPT: Zero-Shot Machine-Generated Text Detection using Probability Curvature & cs.CL & \url{http://arxiv.org/abs/2301.11305v2} & 4 & 76 & 12.481\\
\bottomrule
 \end{tabular}
 \caption{Papers, their prime category, arXiv link, week of first arXiv submission, citation count (as of 07/29/2023) and z-score. \textbf{Top 20 papers} according to z-score among all 
 \arxivlarge{} papers.} 
 \label{table:top20}
 \end{table}

\begin{table}[!htb]
 \centering
 \footnotesize
 \begin{tabular}{|p{0.3cm}|p{5.5cm}|p{1cm}|p{4.5cm}|p{0.7cm}|p{0.5cm}|p{1cm}|}
\toprule
 No. & Title & Cat. & Link & Week & Cit&  z-score \\ \midrule
21 & Mastering Diverse Domains through World Models & cs.AI & \url{http://arxiv.org/abs/2301.04104v1} & 2 & 59 & 12.238\\
22 & Augmented Language Models: a Survey & cs.CL & \url{http://arxiv.org/abs/2302.07842v1} & 7 & 79 & 12.079\\
23 & A Comprehensive Survey on Pretrained Foundation Models: A History from BERT to ChatGPT & cs.AI & \url{http://arxiv.org/abs/2302.09419v3} & 7 & 79 & 12.079\\
24 & ImageBind: One Embedding Space To Bind Them All & cs.CV & \url{http://arxiv.org/abs/2305.05665v2} & 19 & 39 & 11.966\\
25 & Muse: Text-To-Image Generation via Masked Generative Transformers & cs.CV & \url{https://arxiv.org/abs/2301.00704} & 1 & 111 & 11.692\\
26 & T2I-Adapter: Learning Adapters to Dig out More Controllable Ability for Text-to-Image Diffusion Models & cs.CV & \url{http://arxiv.org/abs/2302.08453v2} & 7 & 76 & 11.609\\
27 & Is ChatGPT a General-Purpose Natural Language Processing Task Solver? & cs.CL & \url{http://arxiv.org/abs/2302.06476v2} & 6 & 145 & 11.328\\
28 & SemEval-2023 Task 2: Fine-grained Multilingual Named Entity Recognition (MultiCoNER 2) & cs.CL & \url{http://arxiv.org/abs/2305.06586v2} & 19 & 36 & 11.024\\
29 & Mathematical Capabilities of ChatGPT & cs.LG & \url{http://arxiv.org/abs/2301.13867v2} & 5 & 79 & 11.016\\
30 & The Flan Collection: Designing Data and Methods for Effective Instruction Tuning & cs.AI & \url{http://arxiv.org/abs/2301.13688v2} & 5 & 78 & 10.873\\
31 & The False Promise of Imitating Proprietary LLMs & cs.CL & \url{http://arxiv.org/abs/2305.15717v1} & 21 & 16 & 10.480\\
32 & The RefinedWeb Dataset for Falcon LLM: Outperforming Curated Corpora with Web Data, and Web Data Only & cs.CL & \url{http://arxiv.org/abs/2306.01116v1} & 22 & 11 & 10.421\\
33 & Distilling Step-by-Step! Outperforming Larger Language Models with Less Training Data and Smaller Model Sizes & cs.CL & \url{http://arxiv.org/abs/2305.02301v2} & 18 & 26 & 10.387\\
34 & Video-LLaMA: An Instruction-tuned Audio-Visual Language Model for Video Understanding & cs.CL & \url{http://arxiv.org/abs/2306.02858v3} & 23 & 12 & 10.136\\
35 & InstructBLIP: Towards General-purpose Vision-Language Models with Instruction Tuning & cs.CV & \url{http://arxiv.org/abs/2305.06500v2} & 19 & 33 & 10.083\\
36 & PandaGPT: One Model To Instruction-Follow Them All & cs.CL & \url{http://arxiv.org/abs/2305.16355v1} & 21 & 15 & 9.804\\
37 & ChatGPT is not all you need. A State of the Art Review of large Generative AI models & cs.LG & \url{http://arxiv.org/abs/2301.04655v1} & 2 & 46 & 9.459\\
38 & Theory of Mind May Have Spontaneously Emerged in Large Language Models & cs.CL & \url{http://arxiv.org/abs/2302.02083v3} & 5 & 68 & 9.440\\
39 & mPLUG-Owl: Modularization Empowers Large Language Models with Multimodality & cs.CL & \url{http://arxiv.org/abs/2304.14178v1} & 17 & 34 & 9.377\\
40 & Otter: A Multi-Modal Model with In-Context Instruction Tuning & cs.CV & \url{http://arxiv.org/abs/2305.03726v1} & 18 & 23 & 9.146\\

\bottomrule
 \end{tabular}
 \caption{Papers, their prime category, arXiv link, week of first arXiv submission, citation count (as of 07/29/2023) and z-score. \textbf{Papers 21-40} according to z-score among all 
 \arxivlarge{} papers.}
 \label{table:top40}
 \end{table}

\section{Top $N$ papers}\label{sec:toppapers}

Table \ref{table:top20} showcases the top 20 papers 
extracted according to the methodology described in Section \ref{sec:methodology}. We make several interesting observations: 


\begin{itemize}
\item 13 out of 20 (65\%) of papers have cs.CL as their prime arXiv category (note that authors of papers may wish to indicate as many additional categories as they desire). cs.LG is the prime category 3 times, followed by cs.CV (computer vision; 2 times) and cs.CR (cryptography) and cs.AI (1 time each). 
\item The absolute citation counts vary drastically, with 14 as lowest number in our top-20 list for a paper published in very late May (\emph{Large Language Models are not Fair
Evaluators} \cite{wang2023large}) and 874 as highest numbers for the LLaMA paper \cite{touvron2023llama} published in late February. The relative citation counts vary from 12 standard deviations above the mean to 28 standard deviations above the mean. 
\item The four dominating papers can be seen as technical reports on 
\textbf{LLM foundations models}, including LLaMA \cite{touvron2023llama} (the paper with the highest z-score), PaLM 2 \cite{anil2023palm}, and GPT4 (represented twice; once as an OpenAI publication without dedicated authors focusing on technical details \cite{openai2023gpt4} and once by a group of Microsoft researchers focused on extensive evaluation \cite{bubeck2023sparks}, both published at around the same time). A ``Survey of Large Language Models'' \cite{zhao2023survey} (rank 10 in our list) published in late March and already updated 11 times further indicates the popularity of diverse LLMs. 
\item While not all being technical reports or surveys, the vast majority of top papers are centered around LLMs (at least 18 out of 20, i.e., 90\%). Exceptions are two papers from the computer vision domain (ranks 7 and 13). 
\item It is interesting that LLaMA \cite{touvron2023llama}, a set of \textbf{efficient} (and open-source) foundation language models, dominates overall. This hints at the importance of efficiency for LLMs in general, both from an environmental perspective but possibly even more so from a practical perspective, as the LLaMA models can still be fine-tuned even by researchers with a `modest' GPU endowment \cite{lee2023surveying}. Efficiency is further represented by QLoRA \cite{dettmers2023qlora}, submitted to arXiv in late May, which discusses efficient fine-tuning of quantized LLMs. 
\item Three top papers \cite{bang2023multitask,guo2023close,shen2023hugginggpt} (ranks 9, 15 and 18) are specifically centered around \textbf{ChatGPT} (arguably as the originator of the new LLM hype \cite{Leiter2023ChatGPTAM}) and particularly discuss its \emph{evaluation} including failure cases.  The paper \cite{shen2023hugginggpt} uses ChatGPT to solve AI tasks by querying huggingface.  
\item Two further top papers (ranks 12 and 14) explore \textbf{problem solving with LLMs}, one using external tools \cite{schick2023toolformer} and one using reasoning strategies \cite{yao2023tree}. 
\item Using \textbf{LLMs for evaluation} is discussed in the two papers \cite{wang2023large,zheng2023judging} (ranks 8 and 17), one for evaluating open-ended dialogue and one discussing biases of evaluation with LLMs. Both papers are much more recent, being published in late May and early June.
\item Two papers \cite{driess2023palme,wang2023voyager} (ranks 5 and 13) discuss \textbf{embodied agents} that can interact with the real world, making use of LLMs.
\item Two papers \cite{kirchenbauer2023watermark,mitchell2023detectgpt} (ranks 19 and 20) can be seen as particularly discussing the \textbf{ethical aspects} of detecting LLM generated text (e.g., for spotting misleading AI generated content or to detect cheating in educational contexts) and watermarking AI generated text, i.e., embedding signals in automatically generated text that allow its algorithmic detection. Both papers were published early on, in late January.  
\item Finally, the exceptions in our top 20 list are two computer vision papers. The \emph{Segment Anything} paper \cite{kirillov2023segment} by Meta AI Research provides a dataset for image segmentation. The paper \cite{carlini2023extracting} discusses privacy of image diffusion models such as DALL-E 2 (which can be considered the analogues of LLMs in the computer vision domain). A further computer vision paper introduces a multimodal framework called LLaVA \cite{liu2023visual}, building on top of GPT4.
\item Recently, there has been a debate whether AI/NLP has become more negative, i.e., whether papers tend to report more negatively regarding ongoing research (e.g., outline limitations and failure cases) \cite{bowman-2022-dangers,DidAIGetMoreNegativeRecently}. In our top-20 list, only two papers (10\%) could be considered critique papers, namely \cite{wang2023large}, which focuses on and uncovers biases in LLMs as evaluation models, and \cite{carlini2023extracting}, which criticizes lack of privacy of diffusion models, allowing to retrieve private information from the training data. In the top-40 list, there are two additional negative papers, i.e., \cite{frieder2023mathematical} which disputes the mathematical capabilities of ChatGPT, and \cite{gudibande2023false}, which challenges whether distillation in which a smaller student LLM is trained on the outputs of a larger properietary LLM such as ChatGPT is really effective. A few papers are partly negative, highlighting some limitations, such as \cite{bang2023multitask}. Overall, the most popular papers are (currently) thus positive regarding the development and abilities of recent LLMs.  
\end{itemize}


Table \ref{table:top40} gives analogous papers with rank 21 to 40. We refrain from an in-depth analysis as above. The papers have a similar scope, however, with 11 out of 20 (55\%) having cs.CL as primary category and 13 out of 20 (65\%) having a variant of LLM in their title (language models, ChatGPT, GPT, etc.). Interestingly, the list of papers with ranks 21-40 contain quite a few \textbf{multimodal} approaches such as text-to-image generation models, and relatively more so than the list of papers with ranks 1-20.


\section{Analysis}\label{sec:analysis}

We now briefly perform a few further analyses on our corpus (not only \arxivsmall{} but also \arxivlarge) in order to better understand recent developments. 

\begin{figure}[!htb]
\centering
\includegraphics[width=\textwidth]{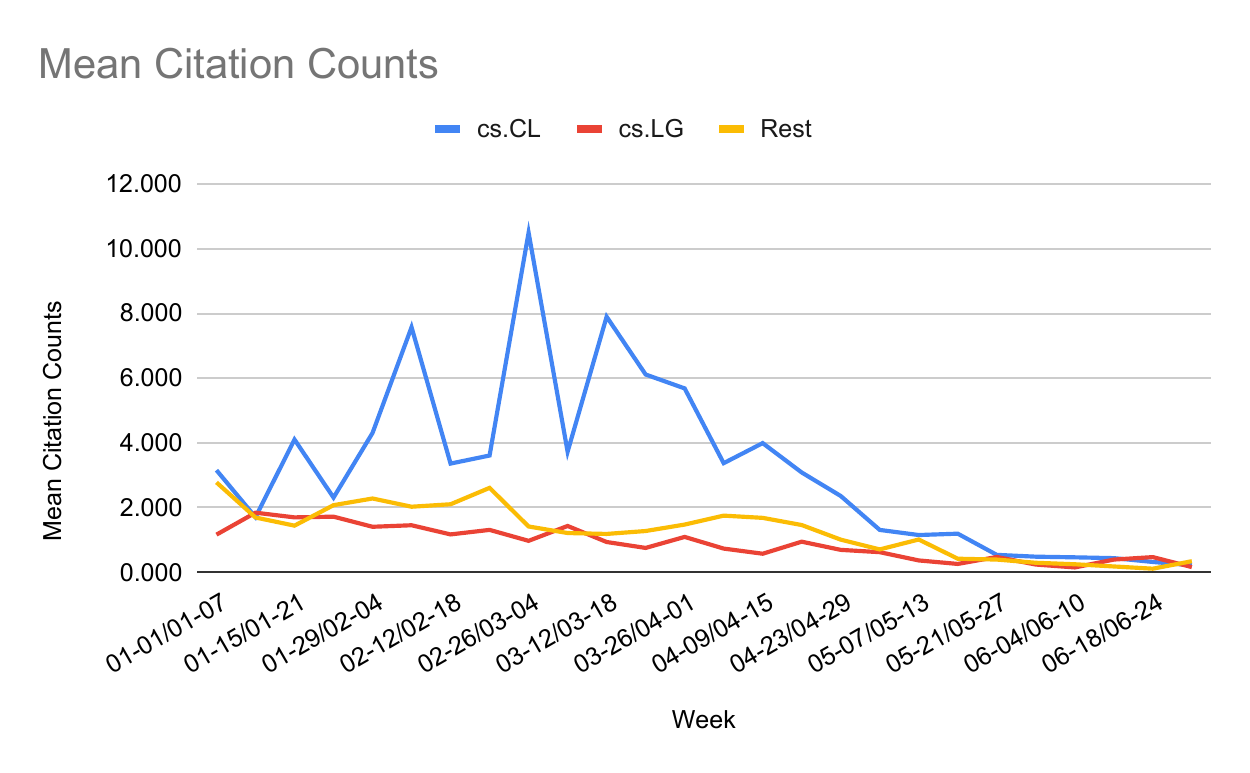}
\caption{Mean number of citations over weeks for different arXiv categories.}
\label{fig:cits}
\end{figure}

\paragraph{How many citations and standard deviations are there per week?} 
Figure \ref{fig:cits} gives the mean citation counts of  papers  
belonging to three primary categories (cs.CL, cs.LG, and all others) over time. We observe that: 
\begin{itemize}
\item citations tend to decrease over time (as is expected; more recent papers cannot yet have been cited so frequently), with, on average, 
decisively fewer than 2 citations per paper starting from May for all three arXiv categories
\item cs.CL attracts (considerably) more citations than 
cs.LG and the aggregation of all other involved primary categories
\item February has been the month with the most impactful papers in cs.CL, especially week 6 (e.g., Toolformer \cite{schick2023toolformer} and ChatGPT analysis \cite{bang2023multitask} submitted to arXiv) and week 9 (e.g., LLaMA \cite{touvron2023llama} submitted)
\end{itemize}
Detailed results including overall standard deviations are also give in Table \ref{table:mean}. Standard deviations are particularly large in weeks 1, 6, 8-13. 

\begin{table}[]
    \centering
    \begin{tabular}{ccccccc} 
    \toprule
         \textbf{Week Number} & \textbf{Week Date}	& \textbf{Mean} & 	\textbf{Std}	& \textbf{Mean cs.CL}	& \textbf{Mean cs.LG} &	\textbf{Mean Rest} \\ \midrule
         1&	01-01/01-07&	2.208&	9.305&	3.164&	1.175&	2.787\\
2&	01-08/01-14&	1.759&	4.677&	1.714&	1.852&	1.703\\
3&	01-15/01-21&	2.029&	6.707&	4.110&	1.707&	1.452\\
4&	01-22/01-28&	1.942&	5.934&	2.318&	1.729&	2.084\\
5&	01-29/02-04&	2.144&	6.977&	4.314&	1.417&	2.290\\
6&	02-05/02-11&	2.647&	12.566&	7.578&	1.464&	2.036\\
7&	02-12/02-18&	1.900&	6.383&	3.368&	1.181&	2.112\\
8&	02-19/02-25&	2.158&	9.813&	3.617&	1.321&	2.618\\
9&	02-26/03-04&	2.651&	31.063&	10.496&	0.983&	1.426\\
10&	03-05/03-11&	1.688&	7.647&	3.734&	1.442&	1.224\\
11&	03-12/03-18&	2.191&	19.967&	7.901&	0.947&	1.196\\
12&	03-19/03-25&	1.930&	14.487&	6.117&	0.764&	1.286\\
13&	03-26/04-01&	2.197&	10.052&	5.688&	1.104&	1.485\\
14&	04-02/04-08&	1.743&	8.802&	3.380&	0.744&	1.761\\
15&	04-09/04-15&	1.680&	6.175&	4.000&	0.585&	1.692\\
16&	04-16/04-22&	1.692&	5.715&	3.093&	0.957&	1.471\\
17&	04-23/04-29&	1.174&	3.501&	2.368&	0.704&	1.020\\
18&	04-30/05-06&	0.885&	2.418&	1.320&	0.633&	0.718\\
19&	05-07/05-13&	0.869&	3.187&	1.162&	0.378&	1.024\\
20&	05-14/05-20&	0.641&	3.231&	1.204&	0.270&	0.431\\
21&	05-21/05-27&	0.496&	1.479&	0.550&	0.485&	0.405\\
22&	05-28/06-03&	0.334&	1.023&	0.494&	0.251&	0.306\\
23&	06-04/06-10&	0.276&	1.157&	0.474&	0.162&	0.260\\
24&	06-11/06-17&	0.338&	1.674&	0.449&	0.405&	0.192\\
25&	06-18/06-24&	0.309&	3.861&	0.331&	0.481&	0.123\\
26&	06-25/07-01&	0.257&	2.163&	0.228&	0.173&	0.354\\
\bottomrule
    \end{tabular}
    \caption{Mean number of citations, over all papers including standard deviations, and for the primary categories cs.CL, cs.LG and the remaining categories.}
    \label{table:mean}
\end{table}

\paragraph{How many arXiv categories (scientific subfields) are involved?}
Our dataset \arxivlarge{} comprises 20,843 papers submitted to arXiv between 01/01/2023 and 06/31/2023 with at least one of the indicated categories given as cs.CL or cs.LG. As NLP and ML affect all aspects of life nowadays, we would expect 
that these papers do not only originate from either ML or NLP. 
Indeed, we find that 
our 20,843 papers are assigned to 123 different primary arXiv categories. 
We give detailed statistics on those 19 primary categories occurring at least 100 times in Table \ref{table:categories}. 
Overall, the most frequent 19 primary categories are made up of 5 top level categories, namely: cs (computer science), stat (statistics), eess (electrical engineering and systems science), math (mathematics) and quant-ph (quantum physics).\footnote{ArXiv does unfortunately not include the humanities or social sciences directly.}  
The five most frequent fine-grained categories are cs.LG, cs.CL, cs.CV (computer vision), stat.ML (statistics, machine learning) and cs.AI (artificial intelligence). 

\begin{figure}[!htb]
    \centering
    \includegraphics[width=\textwidth]{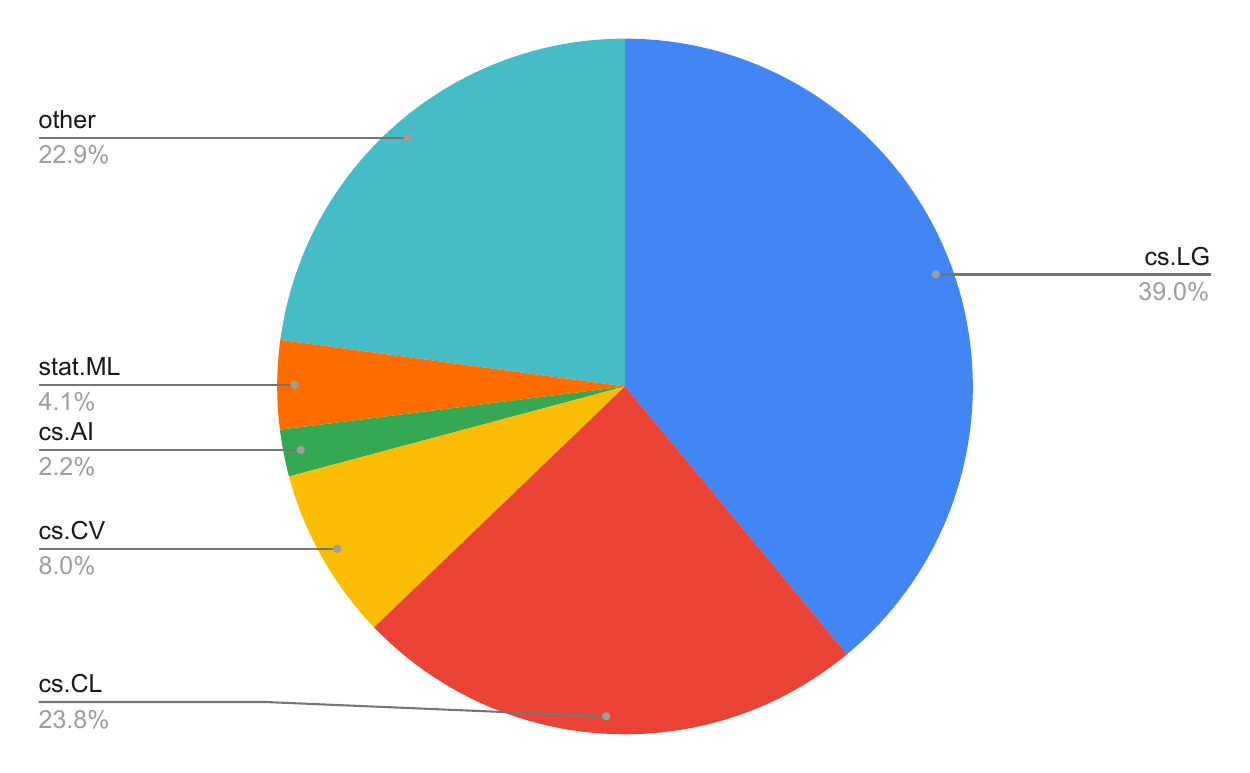}
    \caption{Pie chart of distribution of main categories in our dataset.}
    \label{fig:categories}
\end{figure}

A pie chart of the distribution of primary categories is shown in Figure \ref{fig:categories}. cs.LG is the largest category, almost 40\% of papers have it as its primary category. cs.CL is only about half the size (but dominates the top-40 papers as discussed above). Other primary categories (outside of the top 5 categories) are about the same size as cs.CL. 

\begin{table}[]
\small
    \centering
    \begin{tabular}{c|r}
\toprule
\textbf{Category} & \textbf{Occurrences} \\ \midrule
cs.LG & 8127\\
cs.CL & 4966\\
cs.CV & 1670\\
stat.ML & 859\\
cs.AI & 455\\
eess.IV & 414\\
cs.CR & 304\\
cs.IR & 288\\
cs.RO & 285\\
cs.SD & 265\\
math.OC & 214\\
eess.AS & 212\\
eess.SP & 201\\
cs.HC & 148\\
cs.NE & 143\\
eess.SY & 134\\
cs.SE & 127\\
quant-ph &125\\
cs.CY &111 \\
\bottomrule
    \end{tabular}
    \caption{All primary categories given in our arXiv dataset whose occurrence exceeds 100. ArXiv categories are described here: \url{https://arxiv.org/category_taxonomy}.}
    \label{table:categories}
\end{table}

\paragraph{What distinguishes top papers from other papers?}
We use the tool of \cite{gao-etal-2019-rebuttal} based on the log-likelihood ratio test \cite{dunning-1993-accurate} to determine unusually frequent words in our top-40 papers \arxivsmall{} vs.\ all other papers. Among the top-10 most distinctive unigrams are \emph{chatgpt, gpt-4, modalities, visual, zero-shot}. Among the top bigrams are \emph{language models, large language, models (llms), wide range}. The singular most important trigram is \emph{large language models}. Conversely, words that characterize papers outside the top-40 the best are jargon referring to an older deep learning era such as \emph{learning, neural, deep, network, neural network, machine learning}, etc. 
While this characterization is very simplistic (it certainly does not satisfy to publish a paper on LLMs to obtain high citation numbers), it is nonetheless insightful. 

Top-40 papers also have way more authors on average (11.8, with a standard deviation of 19.5) compared to the remaining papers (4.5 with a standard deviation of 3.2). Part of the effect could be trivial: more authors can increase self-citation counts (an arguably at least partly unethical practice \cite{seeber2019self}). On the other hand, more fundamental research may require a larger author list and industry may also produce papers with a higher number of authors. 

\begin{figure}[!htb]
    \centering
    \includegraphics[width=\textwidth]{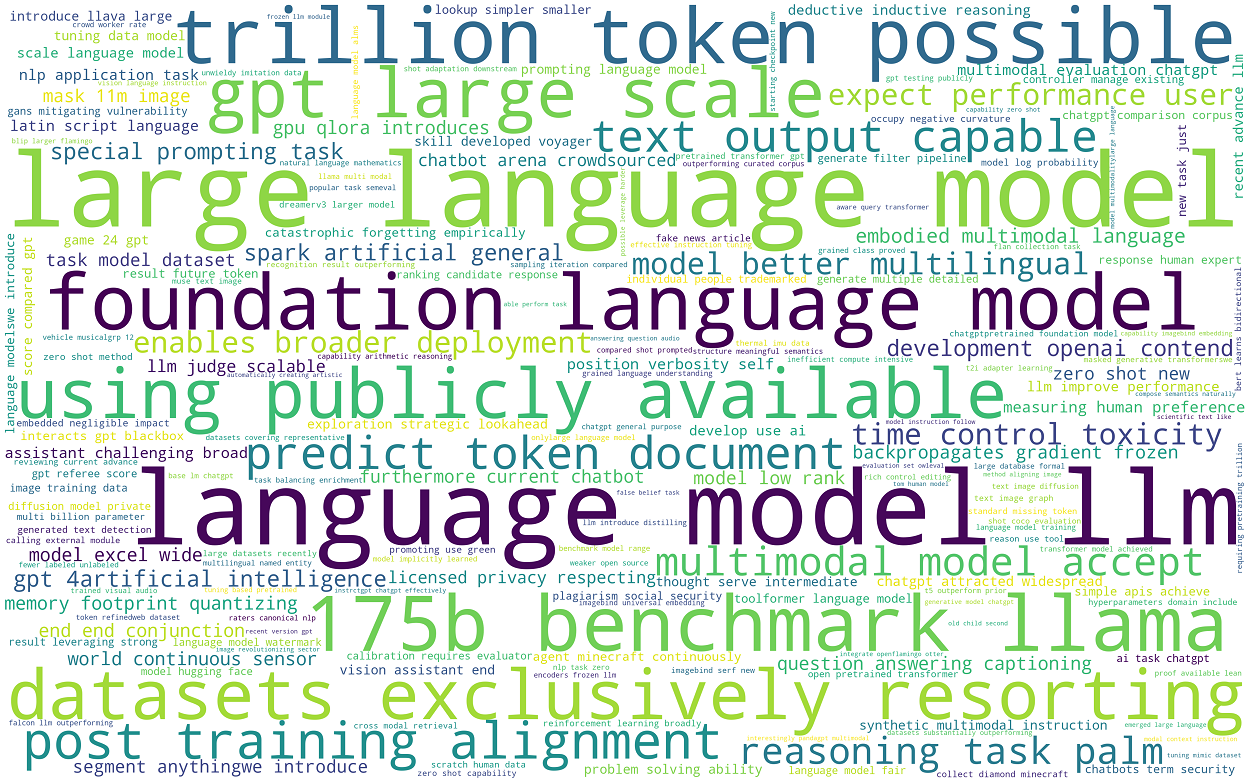}
    \caption{Wordcloud based on the top-40 papers.}
    \label{fig:wordcloud}
\end{figure}

\paragraph{What are the most important key words of the top-40 papers?} We plot a wordcloud of the top-40 papers (see Figure \ref{fig:wordcloud}). To do so, we use KeyBERT \cite{grootendorst2020keybert} to identify the 5 most important tri-grams from the title and abstract of each paper. Then we filter out a manually selected list of unimportant words and lemmatize each word. Finally, we use the python library \textit{wordcloud}\footnote{\url{https://github.com/amueller/word\_cloud}} for plotting. Here the focus of current research into ever larger models becomes apparent again, with phrases such as \textit{trillion token}, \textit{175b}, \textit{large scale} and \textit{large language model}. The keywords \textit{publicly available} also show a focus on non-proprietary data and models.

\paragraph{How popular are LLMs over time in our arXiv dataset?} 
While we have seen that LLMs are the dominating theme in the top-40 paper list, we wonder how the popularity of LLMs and ChatGPT have developed over time in our complete arXiv dataset \arxivlarge. To this end, we query the keywords ``LLMs'' and ``ChatGPT'' in our dataset over time and flag a paper as relevant if it contains the keywords in its title or abstract.\footnote{We lowercase abstracts and titles, and we look for the keywords ``llm(s)'' and ``large language model(s)'' for LLM; for ChatGPT, we look for ``chatgpt'' and ``chat-gpt''.} 

\begin{figure}[!htb]
    \centering
    \includegraphics[width=\textwidth]{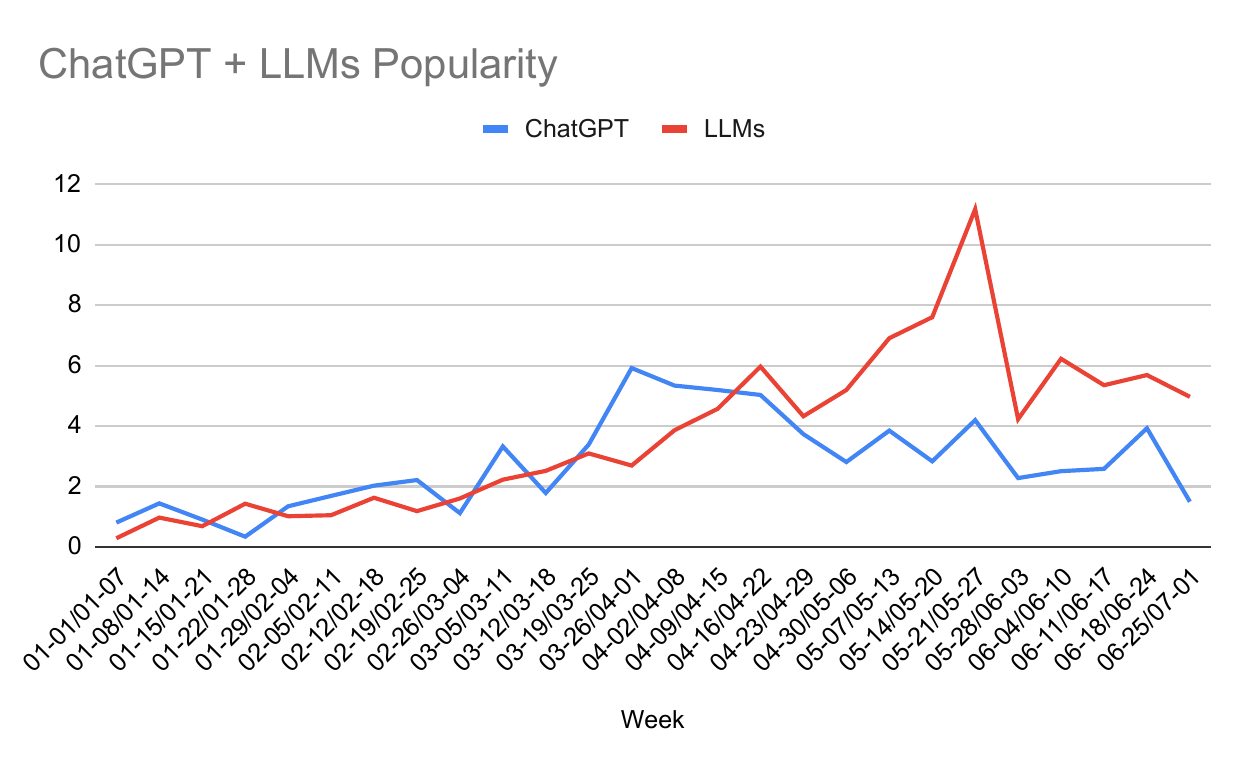}
    \caption{Popularity of ChatGPT and LLMs (in percentage of papers having the words in their abstracts or titles)  over time in our dataset.}
    \label{fig:popularity}
\end{figure}

Figure \ref{fig:popularity} shows the results. 
Both keywords were not very relevant in early 2023, less than 2\% of papers contained them in January. The ChatGPT curve increases until late March (6\% of all papers). Starting from mid-April, LLMs become the more popular keyword. ChatGPT as a keyword declines since then, while LLMs spike in the week of 05/21 (which marks  2023's camera-ready submission deadline for the popular NLP conference ACL \url{https://www.aclweb.org/portal/content/acl-2023-call-papers}) with almost 12\% of papers containing it; we assume that many accepted ACL papers (with LLMs as a topic) were posted to arXiv right after the camera-ready deadline. 
Since then, LLMs seem to be declining as a keyword, also --- even though this could just be an artefact of the conference deadline. 

\section{Conclusion}\label{sec:conclusion}
We have examined arXiv papers related to the categories cs.CL and cs.LG over the first half of 2023. First, we sorted papers according to their normalized citation counts, finding that LLM related papers clearly dominate. Within LLMs, the most popular current issues center around: efficiency, LLM based evaluation, ethical aspects, embodied agents and problem solving with LLMs (only slightly less prominent are multimodal approaches encompassing language and other modalities such as images, with at least 8 papers within the top-40). We have also looked at, among others: (i) what characteristics top papers have relative to papers outside the top-40 list in terms of number of authors and vocabulary, (ii) the distributions of citations in our dataset, and (iii) the popularity of ChatGPT, which `caused' the current hype surrounding LLMs in late 2022, and LLMs over time. We hope that our investigation is beneficial not only to newcomers and outsiders to the field of NLP and ML (of which there are seemingly very many nowadays, given how popular the fields have become \cite{Ziems2023CanLL}), providing quick links to useful starting literature, but also to established researchers and their doctoral students. 

In the future, we want to regularly update the current report to see how tastes shift over time, examine our arXiv datasets \arxivlarge{} and \arxivsmall{} in much more depth, and include further arXiv categories related to AI fields (e.g., cs.CV, stat.ML, cs.AI) into our datasets, among others. 

\section*{Limitations}

Limitations of our approach include the following. First of all, science tools like SemanticScholar or GoogleScholar make quite a few mistakes in correctly attributing citations. While we did not study this in depth, we note for example that LLaMA (our top paper) has 874 citations according to SemanticScholar (July 29, 2023) but only 710 citations according to GoogleScholar, a relative difference of $\frac{164}{874} = 18.7\%$. The paper with fewest citations in our top 20 list \cite{wang2023large} has 14 citations (July 29, 2023) according to SemanticScholar but only 9 citations according to GoogleScholar, a relative difference of $\frac{5}{14} = 35.7\%$. While we do think that our rankings are relatively reliable, such deviations may naturally bias our selection of papers, assumedly with higher uncertainty for low citation papers. Secondly, focusing particularly on highly cited papers may induce a bias towards these papers similar to that of a self-fulfilling prophecy or preferential attachment. Thirdly, our focus on weekly citation averages may have unexpected effects: for example, a younger paper with more citations could be ranked below an older paper with fewer citations, for example, if that older paper was published in a week with fewer average citations (e.g., in the early weeks of January where research, and other human activity, is typically less productive, at least in relevant parts of the world, due to preceding holiday activities). Finally, some authors and research groups, potentially more traditional ones, may refrain from submitting their papers to arXiv, despite its otherwise high popularity particularly in the computer science community (see exponential submission growth rates of arXiv submission numbers in the last decades \url{https://info.arxiv.org/help/stats/2021_by_area/index.html}). Papers from such authors or groups will not be part of our dataset and analysis. 

Our limitations must be kept in mind when interpreting our results. 

\section*{Acknowledgements}
The NLLG group gratefully acknowledges support from the Federal Ministry of Education and Research (BMBF) via the interdisciplinary AI research grant ``Metrics4NLG''. Steffen Eger is further supported by the DFG Heisenberg grant EG 375/5-1. We thank Andreas ``Max Power'' Rücklé for thoughtful discussions. 

\bibliographystyle{plain}
\bibliography{my} 

\end{document}